\documentclass[twocolumn,amsmath,amssymb,10pt,nofootinbib,superscriptaddress]{revtex4}
\usepackage{ graphicx, float,amsmath, amsmath, amssymb}
\usepackage{xcolor} 
\usepackage[export]{adjustbox}
\usepackage[breaklinks, colorlinks, citecolor=blue]{hyperref}
\usepackage[labelsep=period]{caption}
\usepackage[normalem]{ulem}
\usepackage{mathtools}
\usepackage{siunitx}
\usepackage{soul}
\usepackage{physics}
\usepackage{minitoc}

\usepackage{bm}
\usepackage{xcolor}
\usepackage{subfigure}

\definecolor{alizarin}{rgb}{0.82, 0.1, 0.26}

\def\be{\begin{equation}}
\def\ee{\end{equation}}
\def\bea{\begin{eqnarray}}
\def\eea{\end{eqnarray}}
\def\bse{\begin{subequations}}
\def\ese{\end{subequations}}

\begin{document}
\setlength{\parindent}{0cm}

\title{Elementary considerations on possible entropy-driven cosmological evolutions}

\author{Jamy-Jayme Th\'{e}zier}%
\thanks{\href{mailto:jamyjayme.thezier@gmail.com}{jamyjayme.thezier@gmail.com}}
\affiliation{%
Laboratoire de Physique Subatomique et de Cosmologie, Universit\'e Grenoble-Alpes, CNRS/IN2P3\\
53, avenue des Martyrs, 38026 Grenoble cedex, France
}

\author{Aur\'{e}lien Barrau}%
\affiliation{%
Laboratoire de Physique Subatomique et de Cosmologie, Universit\'e Grenoble-Alpes, CNRS/IN2P3\\
53, avenue des Martyrs, 38026 Grenoble cedex, France
}

\author{Killian Martineau}%
\affiliation{%
Laboratoire de Physique Subatomique et de Cosmologie, Universit\'e Grenoble-Alpes, CNRS/IN2P3\\
53, avenue des Martyrs, 38026 Grenoble cedex, France
}




\date{\today}
\begin{abstract} 
For several independent reasons, the idea that notorious sources of entropy could exist in the Universe has been recently revived. Taking advantage of a new framework accounting for non-equilibrium processes in cosmology, we explicitly investigate the cosmological dynamics as a function of the entropy production, focusing on the stability of the system. An exhaustive investigation is performed.
As the main physical conclusion, we show that for a wide class of entropy source terms, the fluid dynamics converges towards an effective cosmological constant. Constraints on the associated entropic force are also obtained.
\end{abstract}
\maketitle

\section{Introduction}

Thermodynamics, especially in its  statistical form, might be the most reliable branch of physics. The fundamental asymmetry between the past and the future -- which does not exist at the microscopic level --, or more precisely {\it oriented causation} that we experience everyday, can always be traced back to the second principle \cite{Rovelli:phi}. All that makes this world non-trivial -- history, memory, evolution -- is grounded in two ingredients: the unavoidable increase of entropy (in both time directions\footnote{This statement often appears as paradoxical although it is quite obvious: as expressed by Boltzmann's $H$-theorem, if no specific boundary condition is imposed, any state of a system has an exponentially higher probability to result from the evolution of a higher entropy state {\it and} to evolve toward a higher entropy state. Consider a cloud of gas: the possibility that the {\it previous} state of the system was a lower entropy one (i.e. a smaller cloud) is obviously deeply suppressed -- this would be an extraordinarily improbable fluctuation -- unless the system was {\it forced} to be into a very unlikely initial state (e.g. in a bottle) by an exterior agent.}) and the smallness of its initial value (see, e.g., \cite{sep-time-thermo,Rovelli:2018vvy} and references therein and \cite{Patel:2017xzn} for a brief introduction). The detailed question to address was recently refined in \cite{Scharnhorst:2024nam}, taking into account subtleties from probability theory. Understanding why the entropy of the early universe was so extraordinarily small,  contradicting the logic of statistical physics itself, which is based on the idea that the larger phase space domain -- for a given coarse graining -- defines the actual macro-state of the system, is a major open question in theoretical physics (see, e.g., \cite{Penrose:1979azm} and references therein). Note that the computation of the total entropy of the observable universe is, in itself, a subtle and highly debated question (see, e.g., \cite{Frampton:2008mw} for a pedagogical introduction, \cite{Egan:2009yy} for additional details and \cite{Profumo:2024hnn} for a recent update).

Lately, very speculative -- but deeply revolutionary -- ideas were suggested in \cite{Rovelli:2014cja,Rovelli:2015dha,Rovelli:2018vvy}. It is basically assumed that entropy is intrinsically relational and that its initial smallness is not a feature of Nature but a property of the macroscopic variables, hence of the coarse graining, selected by the coupling of the specific subsystem we constitute with the Universe. Another radically new view was suggested in \cite{Cortes:2022ljn,Cortes:2022hkd,Cortes:2022xnd} where the phase space of biological systems is assumed {\it not} to be fixed, therefore allowing a huge entropy production -- far beyond what would be expected from a reductionist approach. A possible link between the current acceleration of the Universe and the emergence of biospheres was even suggested. 

Independently of those ``exploratory" ideas, the seminal article of Jacobson \cite{Jacobson:1995ab} -- and the numerous following works (see e.g. \cite{Cribiori:2023ffn} for new ideas and references therein) -- deriving the Einstein equations from a proportionality between entropy and horizon area is obviously a far reaching proposal. This has deeply renewed holographic considerations in cosmology (see \cite{Bousso:2002ju}) and opened the path to a fully thermodynamical understanding of gravitation at the very fundamental level \cite{Padmanabhan:2015zmr,Padmanabhan:2009vy}.\\

In this work, we will not go into the details of such models. Instead, we shall clarify how an unexpected  source of entropy, whatever its origin, can drastically modify the cosmological expansion. Taking advantage of the recent developments suggested in \cite{Espinosa-Portales:2019peb}, we shall investigate the dynamics of the Universe while remaining agnostic about the actual physical process taking place.\\

Although the interplay between thermodynamics and gravity is rich and subtle (see, e.g., \cite{Shahhoseini:2025sgl}, and references therein for connections with experiments), one should distinguish between approaches trying to recover the usual gravitational equations (possibly with small modifications) from statistical physics consideration applied to geometrical degrees of freedom -- like in \cite{Cai:2005ra} and subsequent works -- and models trying to add a non-gravitational entropic source to standard general relativity. This work lies in the second category. It does not aim at claiming that the ontology of gravity is entropic, it tries to explore the consequences of processes generating a large amount of entropy in the framework of Einstein's gravity. As we shall see in the following, this leads to an effective modification of only one of the Friedmann equations, hence deserving a specific investigation.\\

We first review the basic ingredients of the model. Then, the stationary points and linear stability of the system are investigated in details. Our analysis is very exhaustive so as to make the work possibly usable in other contexts. Comparisons with actual data are performed to ensure that the effective equation of state is compatible with measurements. Associated constraints on the entropic force are then derived.
We show that, for ``fast varying" sources of entropy, the system spontaneously evolves towards a de Sitter state.

\section{Modified Friedmann equations}

An elegant way to take into account irreversible processes and non-equilibrium phenomena in cosmology, for arbitrary matter and gravity contents, was developed in \cite{Espinosa-Portales:2019peb,Garcia-Bellido:2021idr,Espinosa-Portales:2021cac,Garcia-Bellido:2024qau}. 
Intuitively, the idea is simple. The first Friedmann equation remains unchanged
\begin{equation}
    H^2=\frac{8\pi G}{3}\rho-\frac{k}{a^2},
    \label{F1}
\end{equation}
whereas the continuity equation is modified so that 
\begin{equation}
    T\frac{dS}{dt}=\frac{d}{dt}(\rho a^3)+p\frac{d}{dt}(a^3),
    \label{cont1}
\end{equation}
to mimic the usual $TdS=dU+pdV$. Standard notations are used: $H$ is the Hubble parameter, $a$ the scale factor, $k$ the spatial curvature,  $T$ the temperature, $p$ the pressure, $S$ the entropy, $t$ the cosmic time, $\rho$ the energy density, $U$ the internal energy, and $V$ the volume. The cosmological constant has not been written since one of the goals of the model is precisely to make it emerge from an entropic source (it can anyway always be absorbed in a $\rho_{vac}$ term with an equation of stat $w=p/\rho=-1$). This leads to a modified Raychaudhuri equation: 
\begin{equation}
    \frac{\ddot{a}}{a}=-\frac{4\pi G}{3}\left( \rho+ 3p-\frac{T\dot{S}}{a^3H}\right).
    \label{ray}
\end{equation}

Any process resulting in an entropy growth generates a cosmic acceleration\footnote{For positive values of the Hubble parameter H.}. Interestingly, as highlighted already in \cite{Espinosa-Portales:2021cac}, this equation looks like the Raychaudhuri equation obtained when considering real fluids with a dissipative pressure (that can be either a bulk viscous pressure \cite{Brevik_2017}, or a pressure associated with particle production processes \cite{Banerjee_2024}, or a combination of the two). The effects of non-equilibrium phenomena on the expansion acceleration can therefore be interpreted as those of an effective viscous pressure, the time-evolution of which being non-trivial as it depends on the way entropy is produced.

One can already notice that this equation might, at first sight, seem to imply a kind of ``self-regulation" of the system: as the entropy increases, the scale factor grows, and its contribution to the dynamics quickly becomes negligible due its $1/a^3$ dependence. We will later show that this reasoning does not always hold.\\

Hopefully, this naive approach to the cosmological role of entropy has been implemented in a rigorous and fully relativistic way \cite{Espinosa-Portales:2021cac} where the Leibniz rule writes
\begin{equation}
\begin{aligned}
    \int d^4x \left(\frac{1}{2\kappa} \frac{\delta (\sqrt{-g}R)}{\delta g^{\mu \nu}} + \frac{ \delta \mathcal{L}_m}{\delta g^{\mu \nu}}\right)\delta g^{\mu \nu} \\
    + \int d^4x  \frac{\partial \mathcal{L}_m}{\partial S} \delta S = 0\,,
\end{aligned}
\end{equation}
$R$ being the Ricci scalar, $g_{\mu \nu}$ the metric, $\mathcal{L}_m$ the matter Lagrangian, and $\kappa=8\pi G$. It is supplemented by the variational constraint due to the second law of thermodynamics:
\begin{equation}
\begin{aligned}
    & \frac{\partial L_{m}}{\partial S} \delta S = \frac{1}{2} F_{\mu \nu} \delta g^{\mu \nu}\,,
\end{aligned}
\end{equation}
where $F_{\mu \nu}$ is the tensorial friction or entropic force. At the end of the day, the modified Raychaudhuri equation (\ref{ray}) is recovered. There are known standard phenomena that can produce sizable amounts of entropy, notably the collapse of stars into black holes, the reheating of the Universe, and phase transitions. As mentioned earlier, more exotic ideas -- like evolving Hilbert/phase spaces -- could even be considered. In the following, we do not specify a particular source but, just the other way around, we  determine the form of the entropy production that would be required to explain the observations.\\

It is immediately clear from Eq. \eqref{ray} that any source of entropy, $\dot{S}>0$, will increase the expansion speed. If the entropic term dominates, the Universe will eventually accelerate. Of course, the scale factor can be arbitrarily renormalized and the dynamics can be fully written in terms of physical variables by noticing that
\begin{equation}
    \frac{T\dot{S}}{a^3H}=\frac{T\dot{s}}{H}+3Ts,
\end{equation}
where $s\equiv S/a^3$ is the entropy density.

\section{Linear stability}
\subsection{Stationary points}

We introduce the effective equation of state,

\begin{equation}\label{effective state parameter}
    w_{\text{eff}} \equiv w - \frac{T\dot{S}}{3 H \rho a^3} ~,   
\end{equation}

where $w$ corresponds to the usual equation of state of the cosmological fluid at equilibrium, such that the continuity equation (\ref{cont1}) can be recast into:

\begin{eqnarray}
\label{conservation equation}
\dot{\rho}&=&-3H(\rho+p)+ \frac{T\dot{S}}{a^3}\nonumber\\
&=& -3H\rho(1+w_{\text{eff}}) \\ \nonumber
&=& -\mathrm{K} \rho^{3/2} (1+w_{\text{eff}})~,
\end{eqnarray}

where $\mathrm{K}\equiv\sqrt{24 \pi G}$.




Assuming a constant value of $w$ and a vanishing spatial curvature, the full dynamics can be described in the $(\rho,w_{\text{eff}})$ plane by the following differential equations:

\begin{equation}\label{système equations}
    \begin{cases}
        \dot{\rho} = - \mathrm{K} \rho^{3/2} (1+w_{\text{eff}})~,\\
        \dot{w}_{\text{eff}}=(w_{\text{eff}}-w) \left[ \frac{\mathrm{K}}{2} \sqrt{\rho} (1+3 w_{\text{eff}}) + \frac{\dot{f}}{f}  \right]~,
    \end{cases}
\end{equation}

in which $f \equiv T\dot{S}$. This system has multiple equilibrium points where both $\dot{w}_{\text{eff}}$ and $\dot{\rho}$ vanish. This deserves a comprehensive study.\\ 

The first point to consider is such that 

\begin{equation}
    X^0 \equiv \left( \rho^0=0,w_{\text{eff}}^0=w \right)~.
\end{equation} 

The second -- and non-trivial -- fixed point $X^*$ has coordinates

\begin{equation}
    X^*=\left( \rho^*=\frac{\dot{f}^2}{\mathrm{K}^2f^2},w_{\text{eff}}^*=-1 \right)~.
\end{equation}

Importantly, this fixed point exists only for positive values of $\dot{f}/f = \dot{T}/T + \ddot{S}/\dot{S}$, since one of its coordinates is obtained from $\mathrm{K} \sqrt{\rho^*} = \dot{f}/f$.\\ 

While $X^0$ does not depend on time, $X^*$ is an explicit function of time (unless $f$ is exponential). In this sense, $X^*$ becomes a moving attractor.
It is important to keep in mind that even if the initial state is at $X^*$, the system will {\it not} remain stationary. If, at a given time, the trajectory reaches $X^*$ then, later on, $\rho^*$ will have changed, leading to a non-vanishing derivative of the effective equation of state. As a consequence, $w_{\text{eff}}$ will move apart from $w_{\text{eff}}^*$, implying a variation of the energy density.\\

There are two other nodes for the system at infinity. Let us consider the change of variable $\xi \coloneqq \rho/(\rho+1)$, leading to $\dot{\rho}=\dot{\xi}/(1-\xi)^2$. This immediately allows to identify the node

\begin{equation}
    X^\infty=\left( \rho^\infty=+\infty,w_{\text{eff}}^\infty=w \right),
\end{equation} 

for which the limit $w_\text{eff}\rightarrow w$ must be taken before $\xi \rightarrow 1$. Another straightforward change of variable of the kind $\alpha \coloneqq w_{\text{eff}}/(w_{\text{eff}}+1)$ leads to:

\begin{equation}
    X^\dagger=\left( \rho^\dagger=0,w_{\text{eff}}^\dagger=+\infty \right)~,
\end{equation} 

where the $\rho \rightarrow 0$ limit must be taken before $\alpha \rightarrow 1$. The nodes $X^0$, $X^\infty$ and $X^\dagger$ do exist for all values of $\dot{f}/f$ and $w$, while the presence of $X^*$ has the additional condition that $\dot{f}/f\geq 0$. 
As we shall see in the following, because of bifurcations, it is important to consider invariants appearing either when $\dot{f}/f=0$ or when $w=-1$. 

If $\dot{f}/f=0$ and $w\neq-1$, the line

\begin{equation}
    L_{w_\text{eff}}=\left\lbrace (\rho,w_\text{eff}) \in \mathbb{R}_+ \times \mathbb{R}~|~ \rho=0 \right\rbrace
\end{equation} 

is an invariant of the set of equations (\ref{système equations}).

If $w=-1$ and $\dot{f}/f \neq 0$ the invariant set is

\begin{equation}
    L_\rho=\left\lbrace (\rho,w_\text{eff}) \in \mathbb{R}_+ \times \mathbb{R}~|~ w_\text{eff}=w=-1 \right\rbrace~.
\end{equation} 

Furthermore, if both conditions $\dot{f}/f=0$ and $w=-1$ are satisfied, the relevant invariant set is $L_{w_\text{eff}} \cup L_\rho$. It is important to emphasize that $X^0$, $X^\infty$, $X^\dagger$, and $X^*$ are still meaningful in these limit cases, except when $\dot{f}/f<0$ and $w=-1$, in which case $X^*$ is not real anymore. In particular, at these bifurcations, $L_\rho$ contains $X^0$, $X^*$ and $X^\infty$ while $L_{w_\text{eff}}$ contains $X^0$, $X^*$ and $X^\dagger$. Nodes and invariant sets corresponding to all the situations considered are summarized in Table \ref{tab:existence points fixes}.

\begin{table}[H]
    \centering
    \begin{tabular}{|c|c|c|c|}
    \hline
         & $\dot{f}/f <0$ & $\dot{f}/f=0$ & $\dot{f}/f >0$ \\
         \hline
        $w>-1$ & $X^0$, $X^\infty$, $X^\dagger$ & $X^\infty$, $L_\rho$ & $X^0$, $X^*$, $X^\infty$, $X^\dagger$\\
        \hline
        $w=-1$ & $L_{w_\text{eff}}$ & $L_{w_\text{eff}} \cup L_\rho$ & $L_{w_\text{eff}}$\\
        \hline
        $w<-1$ & $X^0$, $X^\infty$, $X^\dagger$ & $X^\infty$, $L_\rho$ & $X^0$, $X^*$, $X^\infty$, $X^\dagger$\\
        \hline
    \end{tabular}
    \caption{Nodes and invariant sets for different $w$ and $\dot{f}/f$ configurations.}
    \label{tab:existence points fixes}
\end{table}

\subsection{Stability}



The Jacobian matrix associated with the system of Eqs. (\ref{système equations}) reads

\begin{equation}\label{general jacobian}
\begin{split}
    &J(X)=\\
    &\begin{bmatrix}
        -\frac{3}{2}\mathrm{K}\sqrt{\rho}(1+w_\text{eff})&~  - \mathrm{K}\rho^{3/2}\\
        \frac{\mathrm{K}}{4 \sqrt{\rho}}(w_\text{eff}-w)(1+3w_\text{eff})&~ \frac{\mathrm{K}\sqrt{\rho}}{2}(6 w_\text{eff} -3w+1) + \frac{\dot{f}}{f} \\
    \end{bmatrix}.
    \end{split}
\end{equation}

This leads to a second order characteristic polynomial given by

\begin{equation}\label{polynome caracteristique}
    \begin{split}
        P_X(\lambda)=& \lambda^2 + \lambda \left[ 3w +2 - 3 w_\text{eff} - \frac{2}{\mathrm{K}\sqrt{\rho}} \frac{\dot{f}}{f} \right]\\
        & + \frac{\mathrm{K}^2 \rho}{4} (w_\text{eff}-w)(1+3w_\text{eff})\\
        &- \frac{3\mathrm{K}^2 \rho}{4}(1+w_\text{eff})(1+6w_\text{eff}+\frac{2}{\mathrm{K} \sqrt{\rho}} \frac{\dot{f}}{f}-3w).
    \end{split}
\end{equation}

Solving the polynomial equation and evaluating at $X^0$ and $X^*$ leads to:

\begin{equation}\label{eigenvalues X^0}
    \begin{cases}
        \lambda^0_1 = 0\\
        \lambda^0_2 = \frac{\dot{f}}{f},
    \end{cases}
\end{equation}

and

\begin{equation}\label{eigeinvalues X^*}
    \begin{cases}
        \lambda^*_+ = - \frac{3 \dot{f}}{4 f} \left[1+w -\sqrt{(1+w)(1/9+w)}\right]\\
        \lambda^*_- =- \frac{3 \dot{f}}{4 f} \left[1+w+\sqrt{(1+w)(1/9+w)}\right].\\
    \end{cases}
\end{equation}

For any point belonging to $L_\rho$, the Jacobian (\ref{general jacobian}) becomes

\begin{equation}\label{jacobian Lrho}
    J(X\in L_\rho)=
    \begin{bmatrix}
        0 & ~-\mathrm{K}\rho^{3/2}\\
        0 & ~\frac{\dot{f}}{f}-\mathrm{K}\sqrt{\rho}\\
    \end{bmatrix},
\end{equation}

with the eigenvalues

\begin{equation}\label{eigenvalues Lrho}
    \begin{cases}
        \lambda_1(X \in L_\rho) = 0\\
        \lambda_2 (X \in L_\rho) =\frac{\dot{f}}{f}-\mathrm{K}\sqrt{\rho}.\\
    \end{cases}
\end{equation}

The extreme cases of $L_{w_\text{eff}}$, $X^\dagger$ and $X^\infty$ are treated in Appendix \ref{appendix:stability}.\\



The physical case, corresponding to $\dot{S} >0$, is obtained for effective equations of state such that $w_\text{eff}<w$. In the following, initial conditions are set accordingly. Importantly, if the inequality $w_\text{eff}<w$ is initially satisfied, it will hold at any time as the $w_\text{eff}=w$ line is never crossed throughout the dynamics. More details regarding this statement are provided in appendix \ref{appendix:stability}. This annex also contains a generic and complete study of the stability, including for example $w_\text{eff}>w$ cases, so that the analysis presented in this work can be easily used for any $f$ (not necessarily $T\dot{S}$), as long as the source term entering Eq.(\ref{conservation equation}) is of the form $f(t)/a^3$.\\


When $\dot{f}/f<0$ the situation is simple. The fixed point $X^0$ is the only stable node of the dynamics in the semi-plane $w_\text{eff}<w$ and it remains an attractor as long as the fluid equation of state satisfies $w>-1$. If the entropy production decreases with time, the effective equation of state goes back to the usual value $w$ and the fluid behaves as a cosmological fluid at equilibrium. In the case of a "phantom fluid" with an equation of state smaller than -1 the usual result stating that the energy density increases while the universe expands is recovered and $X^\infty$ now becomes the attractor for the dynamics, reached at finite time.\\

When $\dot{f}/f>0$ the situation becomes more interesting. The node $X^0$ is repulsive and the stability of the other equilibrium point $X^*$ depends on the value of $w$. As long as $w>-1$, $X^*$ is an attractor. More precisely, for $w\geq-1/9$, $X^*$ is a linear attractor. If $-1<w<-1/9$  trajectories around the stable node $X^*$ are convergent spirals. Notably, $w_{\text{eff}}$ exhibits damped oscillations around $w_{\text{eff}}^*=-1$ with a period given by Eq. (\ref{period of oscillations}).\\

When $w=-1$ a bifurcation takes place, which switches the stabilities of $X^*$ and $X^\infty$. The cosmological fluid still behaves like a cosmological constant at late times.\\

\begin{figure}[H]
    \centering
    \includegraphics[width=\linewidth]{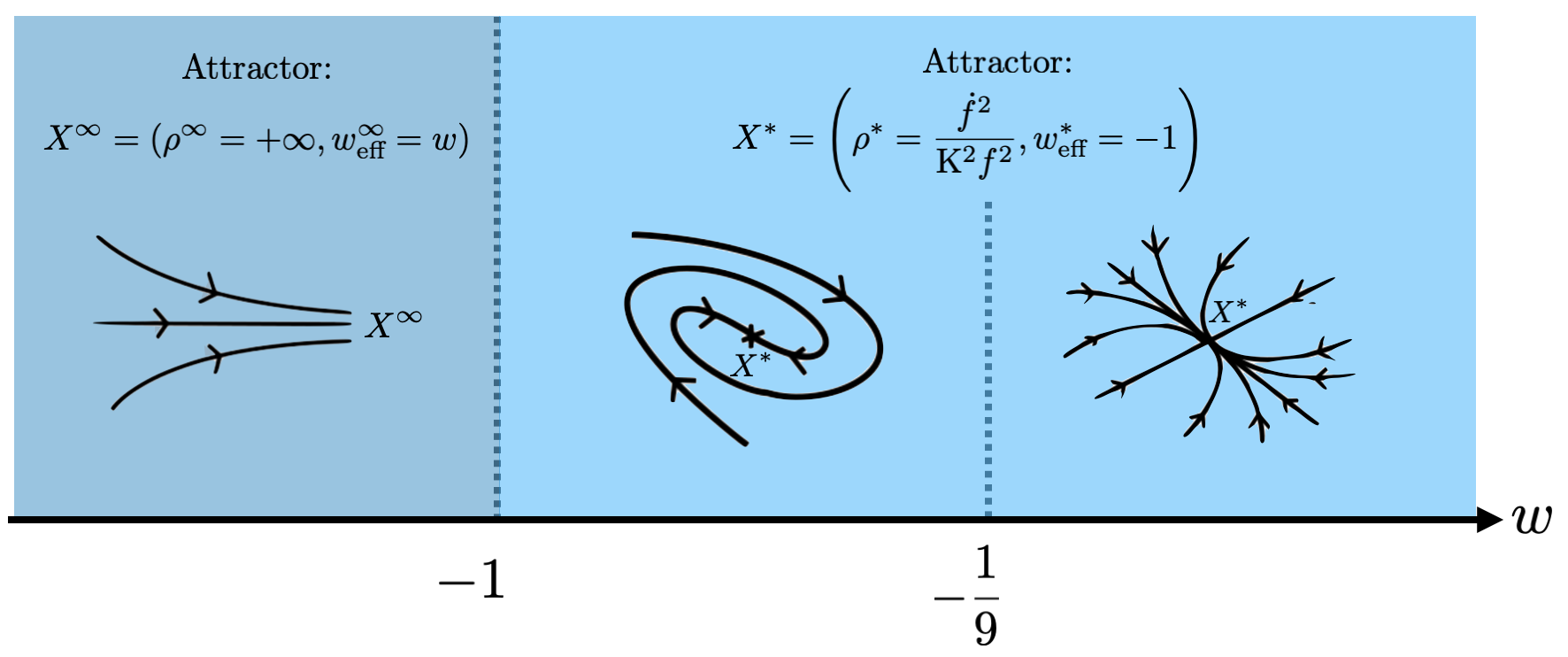}
    \caption{Graphic illustration of the dynamics attractors in the $(\rho, w_{\text{eff}})$ plane as a function of the cosmological fluid equation of state $w>w_\text{eff}$, for positive values of the ratio $\dot{f}/f$.}
    \label{fig:figureAttractors}
\end{figure}

If the fluid equation of state $w$ happens to be smaller than -1, then $\lambda^*_+$ is positive while $\lambda^*_-$ is negative and the node $X^*$ becomes repulsive. The dynamics is, in this case, once again attracted towards $X^\infty$. In particular, the divergence towards $X^\infty$ takes place at a finite time. Figure \ref{fig:figureAttractors} summarizes the situation when $\dot{f}/f>0$, in the plane $w_\text{eff}<w$. This is the important physical picture.\\

The attractor status of $X^*$ is one of the main results of this work. As long as $\dot{f}/f = \dot{T}/T + \ddot{S}/\dot{S} >0$, any cosmological fluid with an equation of state $w \geq -1$ will behave as dark energy at late times. This obviously requires the system entropy to increase fast -- and for long -- but it also shows that the intuition according to which \textit{any} entropy production should play no role at late times is not correct.

\subsection{Matter and radiation}

If the cosmic fluid is matter-dominated, that is if $w=0$, the eigenvalues become

\begin{equation}\label{eigenvalues matter}
    \begin{cases}
        \lambda^0_1 = 0\\
        \lambda^0_2 = \frac{\dot{f}}{f}
    \end{cases}
    \begin{cases}
        \lambda^*_+ = - \frac{\dot{f}}{2f}\\
        \lambda^*_- =-\frac{\dot{f}}{f}\\
    \end{cases}.
\end{equation}

\begin{figure}[H]
    \centering
    \includegraphics[width=\linewidth]{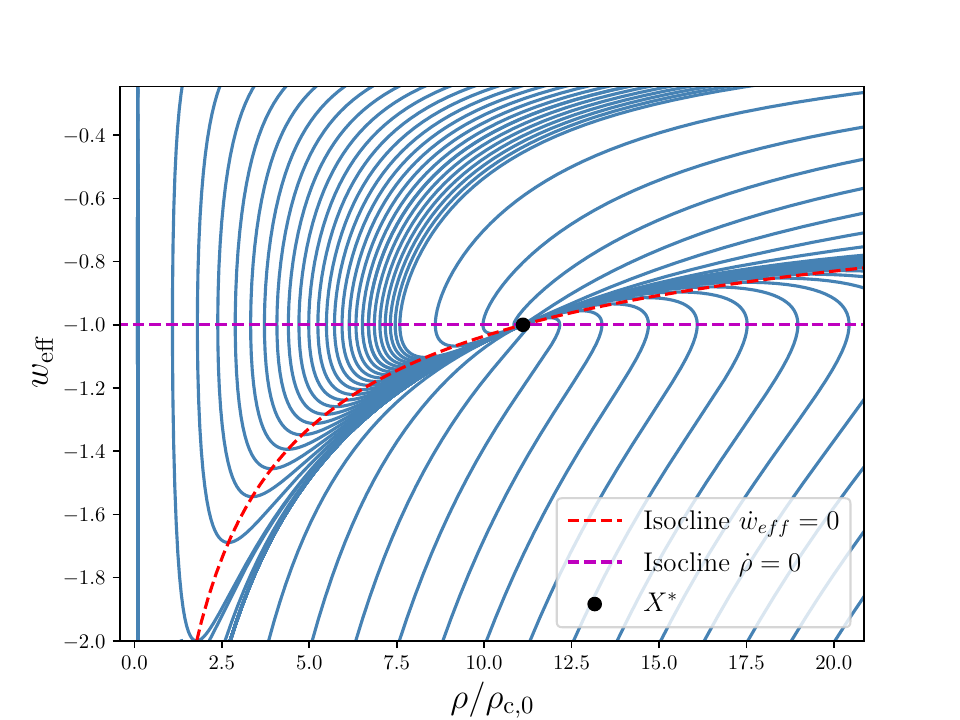}
    \caption{Trajectories in the $(\rho, w_{\text{eff}})$ plane with $\rho_{c,0}=3H_0^2/(8\pi G)$, for $w=0$ and $\dot{f}/f>0$. An exponential $f$ function was chosen, ensuring that $X^*$ (to which trajectories converge) does not evolve with time.} 
    \label{fig:trajectories for w=0 around Xstar}
\end{figure}

For positive values of $\dot{f}/f$ the attractor is $X^*$ (see Fig. \ref{fig:trajectories for w=0 around Xstar}) while it is $X^0$ for $\dot{f}/f<0$. When $\dot{f}/f=0$ a transcritical bifurcation occurs at fixed value of $w$.

In the case of a radiation energy content the equation of state is $w=1/3$ and the Jacobian eigenvalues then read

\begin{equation}\label{eigenvalues radiation}
    \begin{cases}
        \lambda^0_1 = 0\\
        \lambda^0_2 = \frac{\dot{f}}{f}
    \end{cases}
    \begin{cases}
        \lambda^*_+ = - \frac{\dot{f}}{f}(1+1/\sqrt{3})\\
        \lambda^*_- =- \frac{\dot{f}}{f}(1-1/\sqrt{3}).\\
    \end{cases}.
\end{equation}


\section{Phantom dark energy from entropy}

The effective equation of state $w_{\text{eff}}$ previously introduced has no lower bound and can take values below -1, mimicking so-called phantom dark energy fluids. Many trajectories in the $(w_{\text{eff}},\rho)$ plane, as displayed in Fig. \ref{fig:trajectories for w=0 around Xstar}, are indeed crossing this line. For some of those trajectories, $w_{\text{eff}}$ will stay below -1 for an infinite amount of time (as this value is approached from below). 
The question of finding a criterion on the initial conditions for $\left\lbrace w_{\text{eff}},\rho \right\rbrace$ to determine whether a given trajectory starting above $w_{\text{eff}}=-1$ will go below this value -- opening the possibility of occurrence of a {\it Big Rip} -- inevitably comes to mind. To this aim, we focus on a matter dominated universe and we consider an exponential behavior of $f=T\dot{S}$.\\ 

\begin{figure}[H]
    \centering
    \includegraphics[width=\linewidth]{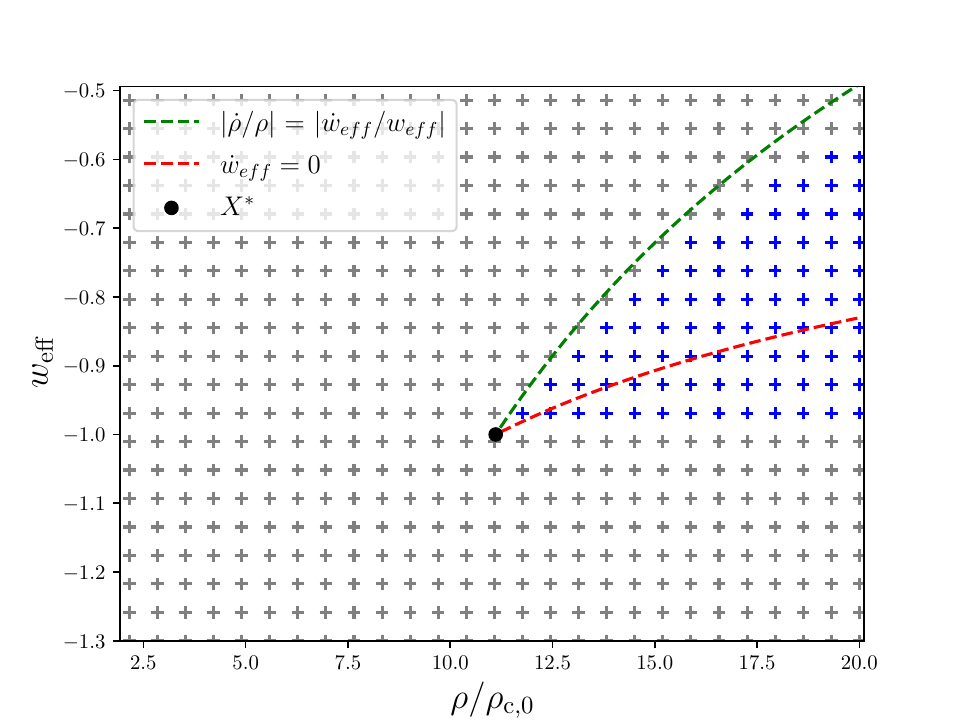}
    \caption{Scan of initial conditions in the case of a matter Universe with an exponential $f$ function. Gray crosses correspond to trajectories that will go below $w_{\text{eff}}=-1$ and blue ones to those always staying above. 
    }
    \label{fig:conditions weff neg}
\end{figure}

Trajectories for which $w_{\text{eff}}<-1$ at some point in the dynamics are represented by grey crosses in Fig. \ref{fig:conditions weff neg}. As can be immediately seen, their set constitute a vast region of the parameter space.

As exhibited on Fig. \ref{fig:trajectories for w=0 around Xstar}, the evolution of any point with coordinates located between $w_{\text{eff}}>-1$ and the isocline $\dot{w}_{\text{eff}}=0$ will tend to stick to this isocline and will approach $ w_{\text{eff}}=-1$ by above.

A precise upper bound on the initial value of $w_{\text{eff}}(\rho)$ such that $ w_{\text{eff}}>-1$ at any time is harder to estimate and we shall just state that it has to be above the isocline:


\begin{equation}\label{condition for the trajectories to stay above -1 lower}
    -\frac{1}{3}(1+2\sqrt{\rho^*/\rho}) \leq w^{\text{up.bound}}_{\text{eff,ini}}(\rho)~.
\end{equation}

\section{Constraints on the entropic force}

We now derive constraints on the values of the first and second derivatives of the current entropy production, under the hypothesis that this accounts for the observed acceleration of the cosmic expansion. We use the standard Chevallier-Polarski-Linder parametrization \cite{CHEVALLIER_2001} for $w_{\text{eff}}$:

\begin{equation}\label{CPL parametrization}
    w_{\text{eff}} (a) = w_0+w_a (1-a).
\end{equation}

The first constraint is straightforwardly obtained by calculating $w_{\text{eff}} (a=1)$ and requiring this to match $w_0$. This gives

\begin{equation}\label{constraint on Sdot}
    \dot{S_0}=\frac{f_0}{T_0}=3 \frac{H_0 \rho_0}{T_0} (w-w_0).
\end{equation}

Numerical values are obtained using $\Omega^M_0\equiv\rho_0 / \rho_c=1$, where $\rho_c$ is the critical density, and $H_0 = h_0 \times 100~ \text{km}.\text{s}^{-1}.\text{Mpc}^{-1}$, with $h_0=0.68$. One should not be surprised by this unusual choice for $\Omega^M_0$. It ensures the correct value of $H_0$ as $\Omega^{\Lambda}_0$ is here assumed to vanish. In this model, $\rho$ is the usual matter density before the beginning of the substantial entropy production. When entropy begins to play a significant role, the dilution of matter slows down, in the sense that it is not anymore proportional to $a^{-3}$. This can be intuitively thought as the excitation of hypothetical growing ``hairs" by the background temperature. Obviously $\rho_0$ does not coincide anymore with the usual estimate of the matter (either dark or visible) content. \\

The second constraint comes from chain rules on $w_a$:

    \begin{align}\label{chain rule wa}
        w_a=\left.-\frac{dw_{\text{eff}}}{da}\right|_{a=1}&=-\left.\frac{dt}{da}\right|_{a=1}\times\left.\frac{dw_{\text{eff}}}{dt}\right|_{t=t_0}\\
        &=-H_0^{-1}\times\left.\frac{dw_{\text{eff}}}{dt}\right|_{t=t_0}.
    \end{align}


With Eq. (\ref{système equations}), this leads to:

\begin{equation}\label{constraint on fdot}
    \dot{f_0}=f_0\left[\frac{H_0 w_a}{w-w_0} - \sqrt{6 \pi G \rho_0} (1+3w_0) \right],
\end{equation}

which can be re-written as a constraint on the second derivative of the entropy:

\begin{equation}\label{constraint on Sdotdot}
    \Ddot{S_0}=\dot{S_0} \left[ \frac{H_0 w_a}{w-w_0} - \sqrt{6 \pi G \rho_0} (1+3w_0) -\frac{\dot{T_0}}{T_0}\right].
\end{equation}

Taking into account observational uncertainties of the relevant quantities (see, {\it e.g.} \cite{Huterer_2017}), one are led with

\begin{equation}\label{f constraints values}
    \begin{cases}
        ( 3.77 \leq f_0\leq 6.26) \times10^{-27} \rm{J.m^{-3}.s^{-1}}\\
        (0.45 \leq \dot{f}_0 \leq 5.84 )\times 10^{-44} \rm{J.m^{-3}.s^{-2}},
    \end{cases}
\end{equation}

which, assuming constant $T_0$, results in:
\begin{equation}\label{entropy constraints values}
    \begin{cases}
        (3.77 \leq \dot{S}_0\leq 6.26)\times10^{-27} \rm{J.T^{-1}_0.m^{-3}.s^{-1}}\\
        (0.45 \leq \ddot{S}_0 \leq 5.84)\times 10^{-44} \rm{J.T^{-1}_0.m^{-3}.s^{-2}}.
    \end{cases}
\end{equation}

\section{Application to the exponential case}

One can easily investigate the behavior of the Universe under the constraints given in Eqs. (\ref{f constraints values}). By construction, it should coincides with the current measurements. Once again, we restrict ourselves to the case where $f(t)=f_0e^{(t-t_0)/\tau}$, which is a straightforward way to model a fast entropy growth starting in the contemporary epoch \cite{biocosmo}.\\

\begin{figure}[H]
    \centering
    \includegraphics[width=\linewidth]{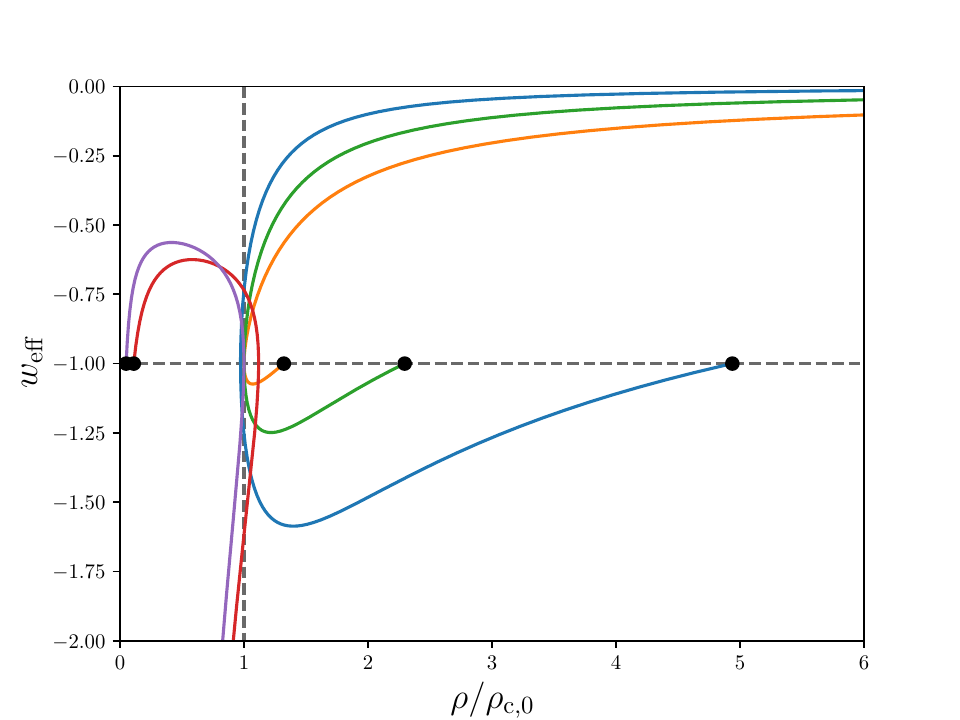}
    \caption{Cosmological evolution in the plane $(\rho/\rho_{c,0},w_{\text{eff}})$. The dark points are the node $X^*$ corresponding to each trajectory (that is the convergence point). For the blue, green and orange lines, time ``starts" in the upper right corner.}
    \label{fig:sec61}
\end{figure}

As shown in Fig. \ref{fig:sec61}, two different behaviors fulfill the constraints of Eqs. (\ref{f constraints values}). 

The first one is usual in the sense that the density diverges in the past, with $w_{\text{eff}}\simeq w=0$. In practice, radiation would, of course, dominate at some point ($w=1/3$) when going backward in time but this is of no relevance here. However, the future is quite different from a simple matter dominated Universe. The density re-increases towards the node $X^*$ and the effective state parameter becomes less than -1. 
This corresponds to blue, green and orange curves on the plot: the final state is the black dot and the current state corresponds to the intersection of the dotted lines. In the remote past, the density increases as the equation of state approaches the one of matter.

A second possible behavior appears when entropy variations are very high -- with a density and an effective equation of state still (marginally) compatible with data. In the past, the entropic term in Eq. (\ref{conservation equation})  and Eq. (\ref{effective state parameter}) dominates over the other terms. 
In such cases, the density goes to zero in the past instead of diverging. The entropic term leads to a positive time derivative of the density in the early Universe, forcing the density to vanish in the remote past.

\begin{figure}[H]
    \centering
    \includegraphics[width=\linewidth]{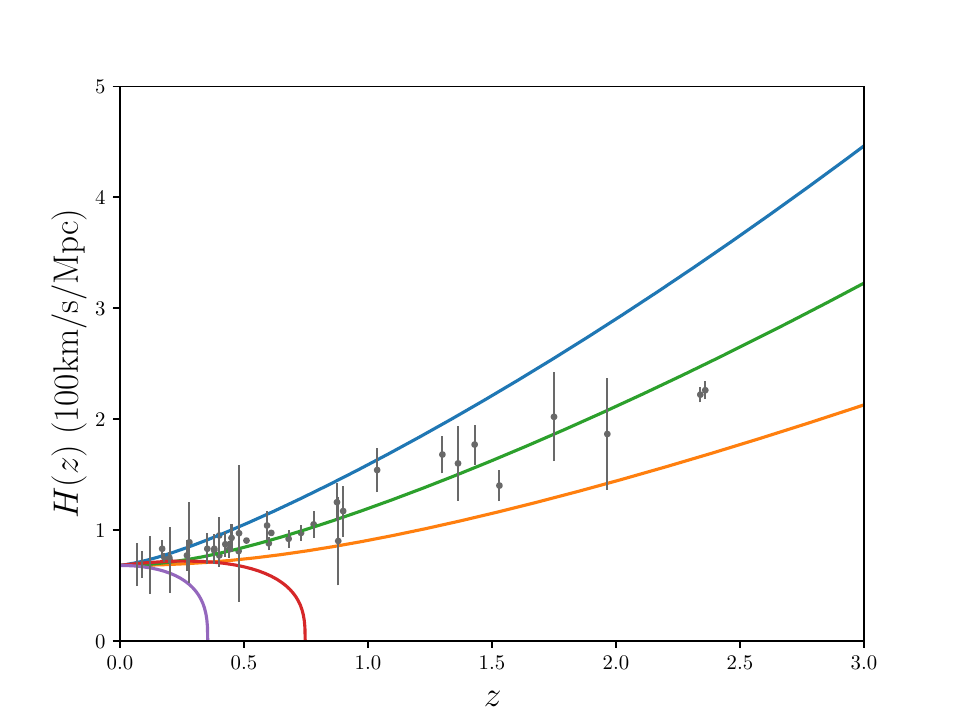}
    \caption{Evolution of the Hubble parameter as a function of the redshift $z$, for the same simulations. Observational data are obtained from \cite{valeursH}.}
    \label{fig:sec62}
\end{figure}

However, going beyond contemporary measurements and using estimations of the Hubble parameter as a function of the redshift (see Fig. \ref{fig:sec62}), one can easily discard this scenario (corresponding to red and purple lines). 

The values of the parameters used in Figs. 4 and 5 are given in Table \ref{tab:valeurs_fig_5}.


\begin{table}[H]
    \centering
    \begin{tabular}{|c|c|c|}
    \hline
        Color & $\tau \; (\rm{H_0^{-1}})$ &  $f_0 \; (\times 10^{-27}\rm{J.m^{-3}.s^{-1}})$ \\
        \hline
        blue  & $0.15$ &  $3.8$\\
        orange & $0.29$ &  $5.2$\\
        green  & $0.22$ &  $5.2$\\
        red  & $1.0$ &  $3.8$\\
        purple  & $1.5$ & $5.2$\\
       \hline
    \end{tabular}
    \caption{Conditions used for $f(t)=f_0e^{(t-t_0)/\tau}$ in the simulations of Figs. \ref{fig:sec61} and \ref{fig:sec62}.}
    \label{tab:valeurs_fig_5}
\end{table}

\section{Conclusion}

In this article, we have investigated the effect of a source of entropy -- assumed to dominate in the Friedmann equations -- on the cosmological dynamics without specifying its nature, therefore keeping the full generality of the approach.\\

The existence and the stability of fixed points has been studied in details. In addition to the study presented in the main corpus, and to make this study useful in different contexts, the dynamics for all possible situations has been exhaustively considered in the Appendix.\\

The main physical result is that the entropic term acts as an effective cosmological constant in the long run, for intense sources of entropy, that is such that $\dot{f}/f>0$.\\

We have taken this opportunity to investigate the possibility that the current acceleration of the cosmic expansion might be due to an entropic term of this kind and we have derived associated bounds.\\

In the future, it would be interesting to investigate in more details what happens when biospheres are considered as sources of entropy, as suggested in \cite{biocosmo,Cortes:2022hkd,Cortes:2022xnd}. 
On the other hand, it would also be interesting to understand in details how the dynamics studied in this work might fit the requirements of primordial inflation, performing a kind of retro engineering: what should the entropy production be to account for CMB data? Can it be explained by standard phenomena? We leave those investigations for future works.

\appendix

\section{Detailed study of the stability}\label{appendix:stability}

For the sake of completeness, we provide here an exhaustive study of the stability of the system of equations (\ref{système equations}). This will allow the use of this work for any physical system described by these equations, in any regime with respect to the bifurcation parameters $\dot{f}/f$ and $w$.\\

The first step to such an analysis is to notice, as mentioned earlier, that the line $w_\text{eff}=w$ is never crossed. We start by providing a short explanation to this statement as it is crucial for the stability analysis.

A trajectory being at some point on this line would have $\dot{w}_\text{eff}=0$, and therefore move only along the density direction. If $w<-1$, then $\dot{\rho} >0$, and the trajectory diverges towards $X^\infty$, while for $w>-1$, $\dot{\rho}<0$, and the trajectory converges to $X^0$. Finally, if $w=-1$, the trajectory does not move at all since $L_\rho$ is invariant throughout the dynamic. The line $w_\text{eff}=w$ thus corresponds exactly to a physical trajectory and the Cauchy-Lipchitz theorem then guarantees that no other trajectory can cross it. This suggests to divide the plane into two parts: one with $w_\text{eff}>w$ and the other with $w_\text{eff}<w$.\\

In the following, we consider the two possible signs of $\dot{f}/f$ independently. In practice, this ratio might, of course, vary and change sign. The behavior of the physical system can however still be understood through this analysis by splitting the time window into smaller ones. For instance, if at the beginning of the trajectory $\dot{f}/f>0$ and $w_\text{eff}<w$ with $w>-1$, the system will converge, as explained below, towards $X^*$. However, if at some point $\dot{f}/f$ changes sign, then the trajectory will turn back towards $X^0$, since $X^0$ is the corresponding attractor.\\

As a final introductory remark, let us stress that the flow diagrams presented in the following
do not represent exact trajectories, but rather show the global trends. In addition, the flow is drawn with a fixed $X^*$ for readability but, as mentioned earlier, this node can move as time goes on. One therefore has to remember that the flow diagram considered may change. It will however keep its key properties as long as $\dot{f}/f$ and $w$ fulfill the specified conditions. 

\subsubsection{Case $\dot{f}/f>0$ and $w>-1$}

In this case, as mentioned earlier, the isocline at $w_\text{eff}=w$ is a trajectory with $\dot{\rho}<0$, thus converging towards $X^0$. This is the only trajectory attracted by $X^0$. The others are repelled and correspond to a positive second eigenvalue of Eqs. (\ref{eigenvalues X^0}).\\

\begin{figure}[H]
    \centering
    \includegraphics[width=\linewidth]{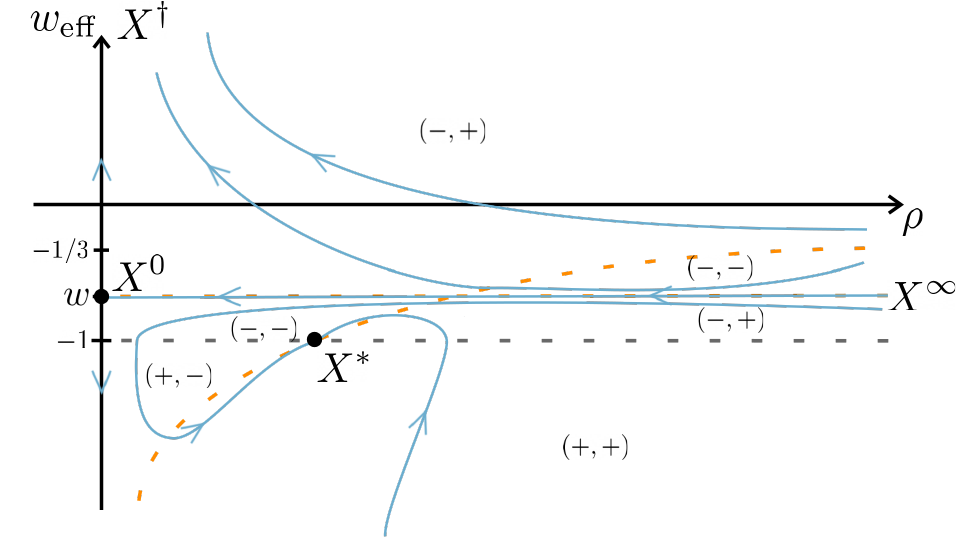}
    \caption{Flow diagram in the case $\dot{f}/f>0$ and $w>-1$ displaying the trajectories (full blue lines), stationary points (blue and black dots, when relevant), the isoclines $\dot{w}_\text{eff}=0$ (orange dashed lines), and the iscoline $\dot{\rho}$ (gray dashed line at $w_\text{eff}=-1$). The nodes $X^\dagger$ and $X^\infty$ are shown to guide the eye, but they are ``at infinity" respectively in the $w_\text{eff}$ and $\rho$ directions. The signs of $\dot{\rho}$ and $\dot{w}_\text{eff}$ are also indicated in regions delimited by the isoclines. For instance $(+,-)$ stands for $\dot{\rho}>0$ and $\dot{w}_\text{eff}<0$.}
    \label{fig:flow diagram ff sup w sup}
\end{figure}

In the semi-plane $w_\text{eff}>w$, the trajectory at $\rho=0$ diverges from $X^0$ towards $X^\dagger$. Taking into account both this information and the sign of the derivatives in the semi-plane (see Fig. \ref{fig:flow diagram ff sup w sup}) leads to the conclusion that $X^\dagger$ attracts all the trajectories in the region $w_\text{eff}>w$.\\

In the other semi-plane, $w_\text{eff}<w$, the eigenvalues of $X^*$, Eqs. (\ref{eigeinvalues X^*}), both have negative real parts. The point $X^*$ is therefore a linear attractor. Furthermore, looking at the sign of the derivatives (see Fig. \ref{fig:flow diagram ff sup w sup}) leads to the conclusion that $X^*$ attracts all the trajectories in the region $w_\text{eff}<w$ (except for the one at $\rho=0$, which diverges towards $w_\text{eff}=-\infty$, which is a marginal case).\\

In the specific case $-1/9>w>-1$, the eigenvalues $\lambda_+^*$ and $\lambda_-^*$ become complex numbers with negative real parts (see Eq. (\ref{eigeinvalues X^*})) and the trajectories around the stable node $X^*$ are convergent spirals. Notably, $w_{\text{eff}}$ exhibits damped oscillations around $w_{\text{eff}}^*=-1$ in the neighbourhood of $X^*$, the period being given by:

\begin{equation}\label{period of oscillations}
    T=\frac{8 \pi}{3} \frac{f}{\dot{f}} \frac{1}{\sqrt{|(1+w)(1/9+w)|}}.
\end{equation}

\subsubsection{Case $\dot{f}/f>0$ and $w=-1$}

Here, $L_\rho$ is an invariant set for the dynamics. The line $w_\text{eff}=w=-1$ is thus made of an infinite number of stationary points. The condition $\lambda_2 (X\in L_\rho)<0$ must be satisfied for the nodes to be stable. Considering the eigenvalues given by Eq. (\ref{eigenvalues Lrho}), this is equivalent to  $\rho>\dot{f}/(\mathrm{K}f)=\rho^*$. Otherwise stated, the node must be under the isocline

\begin{equation}\label{isocline en racine}
    w_\text{eff}=-\left(\frac{1}{3}+\frac{\dot{f}}{f} \frac{2}{3\mathrm{K}\sqrt{\rho}}\right).
\end{equation}

\begin{figure}[H]
    \centering
    \includegraphics[width=\linewidth]{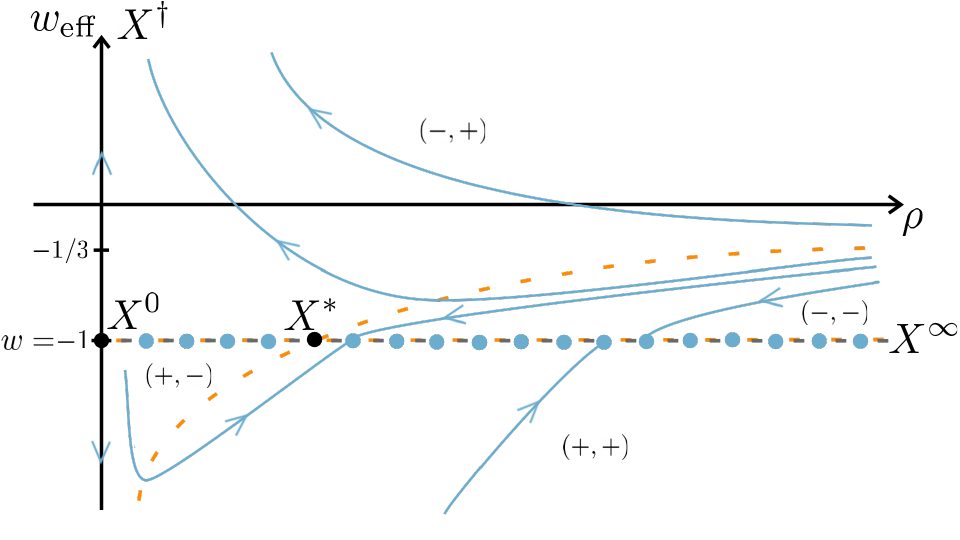}
    \caption{Same as Fig. \ref{fig:flow diagram ff sup w sup} with $\dot{f}/f>0$ and $w=-1$.} 
    \label{fig:flow diagram ff sup w egal}
\end{figure}

This shows that nodes between $X^*$ and $X^0$ are repulsive. Studying the derivatives (see Fig. \ref{fig:flow diagram ff sup w egal}), it is clear that all trajectories such that $w_\text{eff}<w$ converge towards $L_\rho$ in the region $\rho>\rho^*$.\\

For $w_\text{eff}>w$, if the initial conditions are above the limit set by Eq. (\ref{isocline en racine}), the dynamics will diverge towards $X^\dagger$. If they are between the isocline (\ref{isocline en racine}) and $L_\rho$, then the situation is more complicated and depends explicitly on $f$. In general, most trajectories  simply converge towards $L_\rho$, while those close to the isocline (\ref{isocline en racine})  go towards $X^\dagger$.

\subsubsection{Case $\dot{f}/f>0$ and $w<-1$}

For $w<-1$, the eigenvalue $\lambda^*_+$ becomes positive (see Eq. (\ref{eigeinvalues X^*})), implying that $X^*$ is now repulsive. Furthermore, $X^0$ is still repulsive since $\dot{f}/f>0$, which implies a positive eigenvalue for this node (see Eq. (\ref{eigenvalues X^0})).\\

\begin{figure}[H]
    \centering
    \includegraphics[width=\linewidth]{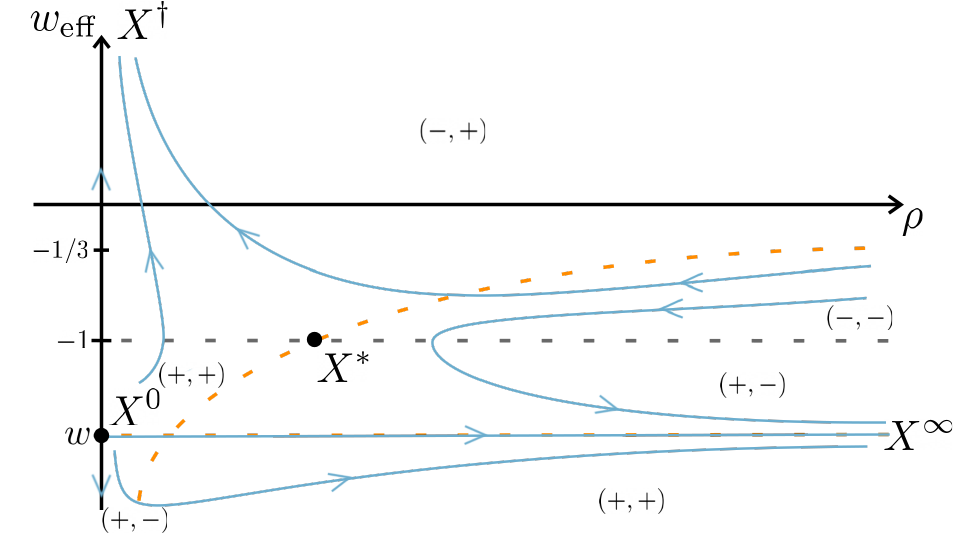}
    \caption{Same as Fig. \ref{fig:flow diagram ff sup w sup} with $\dot{f}/f>0$ and $w<-1$.} 
    \label{fig:flow diagram ff sup w inf}
\end{figure}

The only nodes that might not be repulsive are $X^\dagger$ and $X^\infty$. Considering the lines $\rho=0$ and $w_\text{eff}=w$, and then taking into account the sign of the derivatives (see Fig. \ref{fig:flow diagram ff sup w inf}), leads to the conclusion that all the trajectories under $w_\text{eff}=w$ diverge towards $X^\infty$ ($\rho=0$ being, again, a marginal case).\\

In the semi-plane $w_\text{eff}>w$, all  initial conditions above the isocline (\ref{isocline en racine}) lead to a divergence towards $X^\dagger$ while the ones under both the isoclines (\ref{isocline en racine}) and $w_\text{eff}=-1$ lead to $X^\infty$. As in the previous section, if initial conditions are set above $w_\text{eff}=-1$ and under the isocline (\ref{isocline en racine}), the system has a more complicated behavior. Some trajectories go towards $X^\infty$, while others reach $X^\dagger$ (see Fig. \ref{fig:flow diagram ff sup w inf}).

\subsubsection{Case $\dot{f}/f<0$ and $w>-1$}

For $\dot{f}/f<0$, $X^*$ does not exist and the isocline (\ref{isocline en racine}) becomes

\begin{equation}\label{isocline en racine f/f neg}
    w_\text{eff}=-\left(\frac{1}{3}-\left|\frac{\dot{f}}{f}\right| \frac{2}{3\mathrm{K}\sqrt{\rho}}\right).
\end{equation}

\begin{figure}[H]
    \centering
    \includegraphics[width=\linewidth]{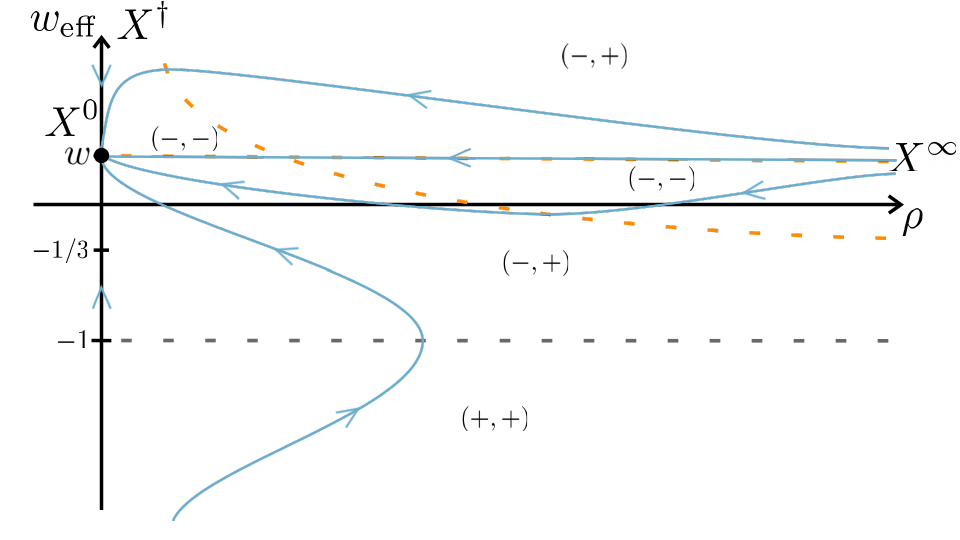}
    \caption{Same as Fig. \ref{fig:flow diagram ff sup w sup} with $\dot{f}/f<0$ and $w>-1$.} 
    \label{fig:flow diagram ff inf w sup}
\end{figure}

This change in the dynamics makes $X^0$ a global attractor. This is straightforward when looking at the signs of the derivatives in Fig. \ref{fig:flow diagram ff inf w sup} and recalling that the eigenvalues (\ref{eigenvalues X^0}) are no longer strictly positive.

\subsubsection{Case $\dot{f}/f<0$ and $w=-1$}

In this case, $L_\rho$ is an invariant subspace. Moreover, the eigenvalue $\lambda_2(X\in L_\rho)$ is negative for any point in the subset. Studying the signs of $\dot{\rho}$ and $\dot{w_\text{eff}}$, one immediately sees that $L_\rho$ is a global attractor for the dynamics.

\begin{figure}[H]
    \centering
    \includegraphics[width=\linewidth]{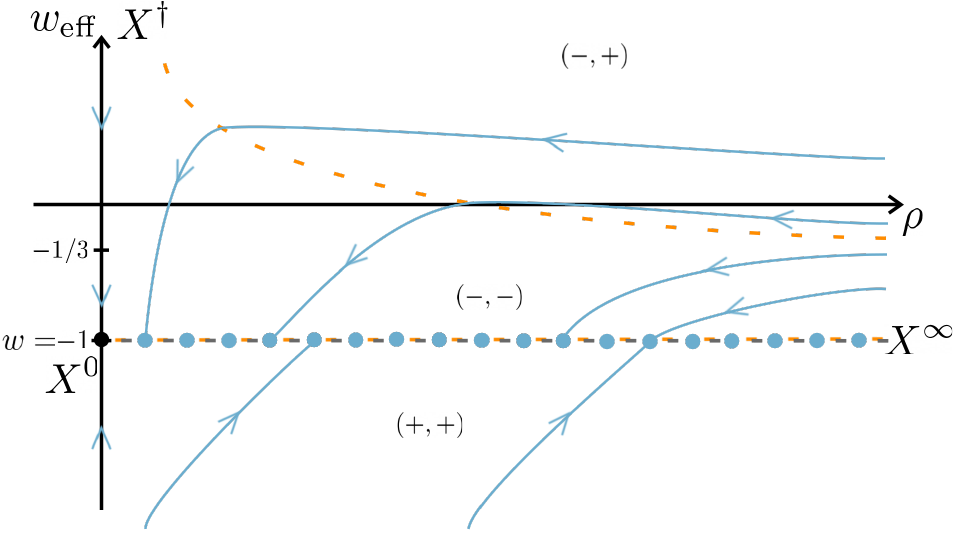}
    \caption{Same as Fig. \ref{fig:flow diagram ff sup w sup} with $\dot{f}/f<0$ and $w=-1$.} 
    \label{fig:flow diagram ff inf w egal}
\end{figure}

\subsubsection{Case $\dot{f}/f<0$ and $w<-1$}

\begin{figure}[H]
    \centering
    \includegraphics[width=\linewidth]{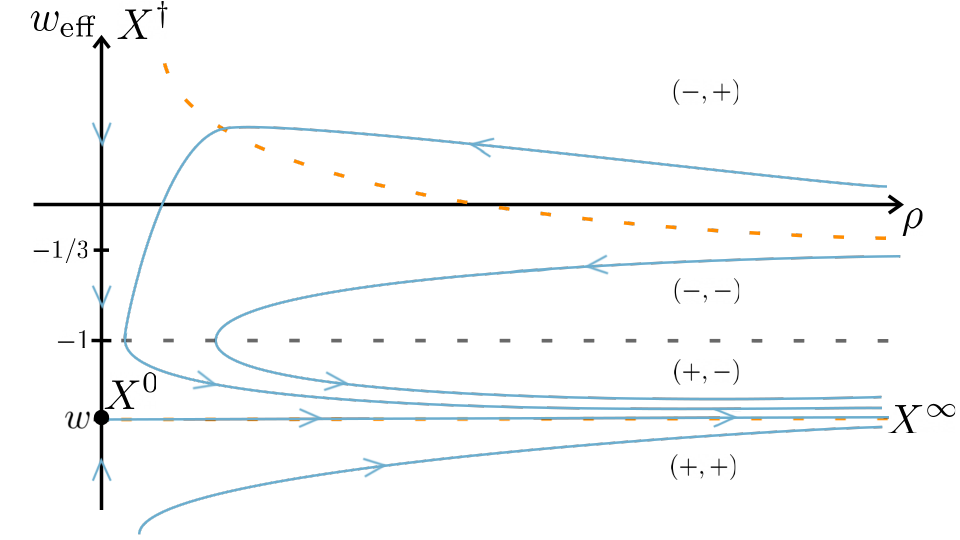}
    \caption{Same as Fig. \ref{fig:flow diagram ff sup w sup} with $\dot{f}/f<0$ and $w<-1$.} 
    \label{fig:flow diagram ff inf w inf}
\end{figure}

In this case, $X^0$ is stable in the direction $w_\text{eff}$. However, it is repulsive in the $\rho$ direction. This is clear by looking at the signs of the derivatives (see Fig. \ref{fig:flow diagram ff inf w inf}). This means that $X^\infty$ is a global attractor for the dynamics (except for the marginal case $\rho=0$).

\subsubsection{Case $\dot{f}/f=0$ and $w>-1$}

For $\dot{f}/f=0$, the isocline (\ref{isocline en racine}) becomes a simple constant,

\begin{equation}\label{isocline en racine f/f=0}
    w_\text{eff}=-\frac{1}{3},
\end{equation}

and corresponds to a trajectory, which cannot be crossed, with a derivative $\dot{\rho}<0$ whatever $w$.\\

Furthermore, the line $\rho=0$ has both $\dot{\rho}=0$ and $\dot{w}_\text{eff}=0$. Consequently, $L_{w_\text{eff}}$ is an invariant set (there is a bifurcation at $\dot{f}/f=0$). It can be seen as a ``ghost" of the transition between the two regimes of isoclines (\ref{isocline en racine}) and (\ref{isocline en racine f/f neg}).

\begin{figure}[H]
    \centering
    \includegraphics[width=\linewidth]{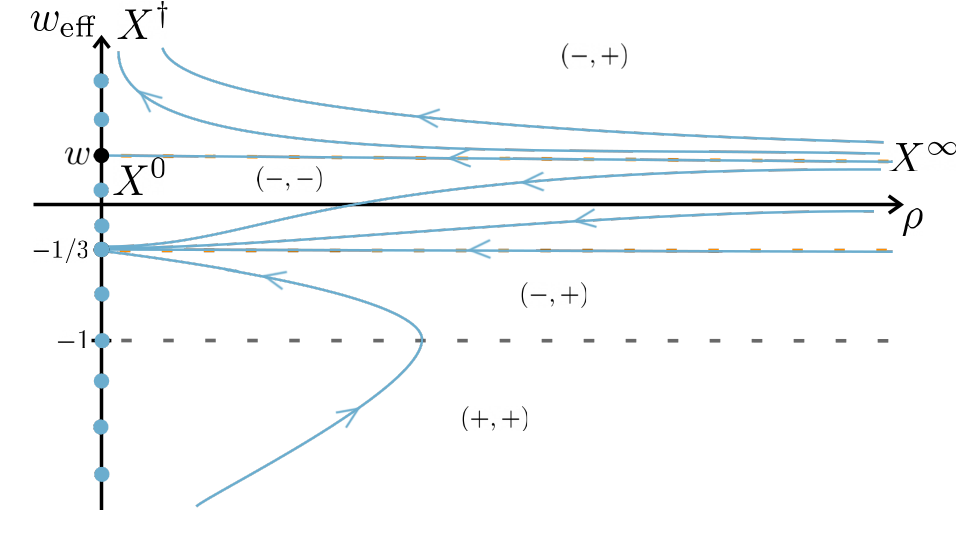}
    \caption{Same as Fig. \ref{fig:flow diagram ff sup w sup} with $\dot{f}/f=0$ and $w>-1/3$. If $-1/3>w>-1$ the stability properties of $X^0$ and $(0,-1/3)$ are swapped.} 
    \label{fig:flow diagram ff egal w sup}
\end{figure}

For $w>-1/3$ and considering as usual the signs of the derivatives, one is led to the flow displayed in Fig. \ref{fig:flow diagram ff egal w sup}. In particular, it should be noticed that $X^\dagger$ attracts all the trajectories above $w_\text{eff}=w$, $X^0$ attracts only the trivial trajectory $w_\text{eff}=w$, and the intersection of the isocline (\ref{isocline en racine f/f=0}) with $\rho=0$, at $(0,-1/3)$, attracts all the other trajectories.\\

It might not seem trivial at first sight that the trajectories ``skip" the nodes of $L_{w_\text{eff}}$ and go towards these attractors. This can be understood by looking at the ratio $\dot{\rho}/\dot{w}_\text{eff}$. Setting the initial condition such that $w_\text{eff}$ is different from $w$ and $-1/3$, leads to a ratio that tends to zero when $\rho$ becomes small, that is near $L_{w_\text{eff}}$. This indicates that the vertical derivative is much larger than the one in the $\rho$ direction, which, intuitively, explains why the nodes of $L_{w_\text{eff}}$ are skipped. This always happens when $\dot{f}/f=0$. Consequently, $L_{w_\text{eff}}$ is unstable except for $(0,-1/3)$ and $X^\dagger$ that are constrained to attract trajectories by the derivatives of the flow.

If $-1/3>w>-1$ then $X^0$ is ``below" the point of coordinates $(0,-1/3)$ in the $(\rho, w_\text{eff})$ plane and the two nodes exchange their stabilities as they both lie on a $\dot{w}_\text{eff}=0$ isocline. This leads to the same shape of the flow as in the previous case (Fig. \ref{fig:flow diagram ff egal w sup}), except that $X^0$ is now the attractor.

\subsubsection{Case $\dot{f}/f=0$ and $w=-1$}

\begin{figure}[H]
    \centering
    \includegraphics[width=\linewidth]{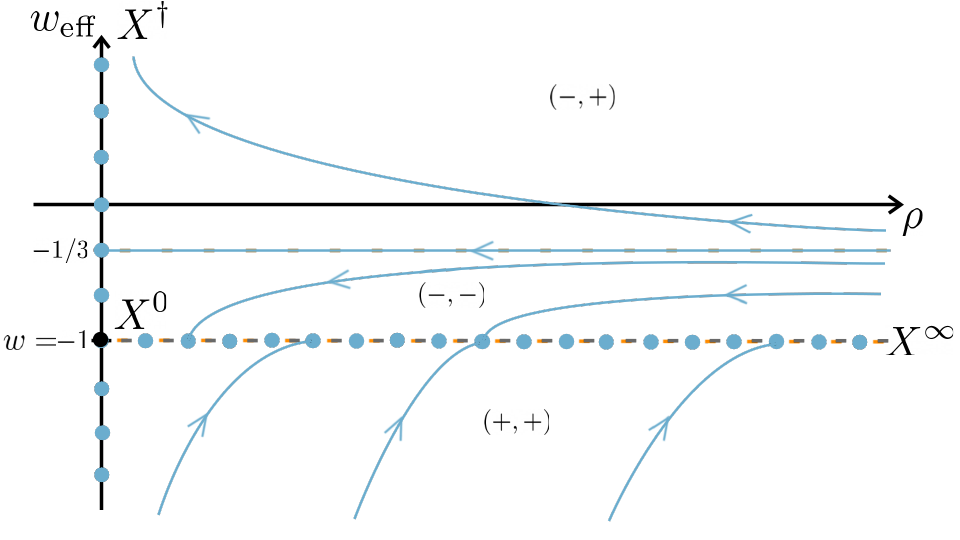}
    \caption{Same as Fig. \ref{fig:flow diagram ff sup w sup} with $\dot{f}/f=0$ and $w=-1$.} 
    \label{fig:flow diagram ff egal w egal}
\end{figure}

In this case, both $L_\rho$ and $L_{w_\text{eff}}$ are relevant. As obvious from in Fig. \ref{fig:flow diagram ff egal w egal}, all the trajectories under the isocline (\ref{isocline en racine f/f=0}) are converging to points of $L_\rho$, while those above diverge towards $X^\dagger$.  The same argument as previously holds to explain why $L_{w_\text{eff}}$ does not attract trajectories but for $X^\dagger$. However, it is slightly more subtle because of $L_\rho$, which is an attractor in this case and actually attracts the flow before it gets to $\rho=0$.

\subsubsection{Case $\dot{f}/f=0$ and $w<-1$}

For this final case, trajectories above the isocline (\ref{isocline en racine f/f=0}) lead to a divergence towards $X^\dagger$ while those under this isocline always diverge towards $X^\infty$. The same argument as previously works to explain why there is no convergence to $L_{w_\text{eff}}$ except at $X^\dagger$.

\begin{figure}[H]
    \centering
    \includegraphics[width=\linewidth]{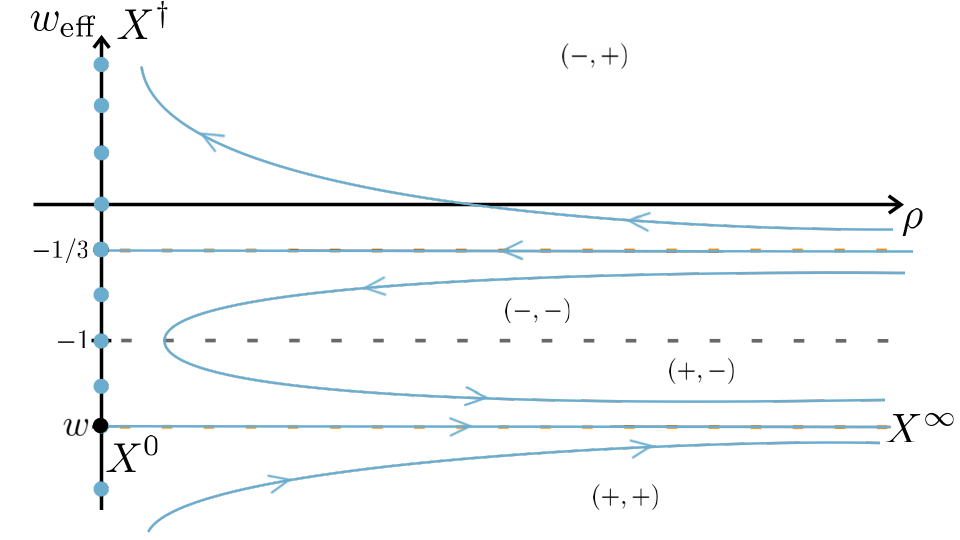}
    \caption{Same as Fig. \ref{fig:flow diagram ff sup w sup} with $\dot{f}/f=0$ and $w<-1$.} 
    \label{fig:flow diagram ff egal w inf}
\end{figure}

\section{Time-Redshift relation}

Most of the quantities measured experimentally are parametrized by the redshift $z$. However, the entropy is usually defined as a function of time.  Deriving a relation between redshift and time in this framework is thus necessary. Let us start from the continuity equation,

\begin{equation}\label{continuity equation without weff}
    \dot{\rho}+3H(1+w)\rho =\frac{T\dot{S}}{a^3}= T (\dot{s}+3Hs),
\end{equation}

where $s=S/a^3$. Solving the differential equation without second member leads to the usual result:

\begin{equation}\label{homogeneous density}
    \rho_H = \rho_{0,Hom} a^{-3(1+w)}.
\end{equation}

Using the variation of parameters, that is injecting $\rho(t)=\rho_{0,Hom}(t) a(t)^{-3(1+w)}$ in Eq. (\ref{continuity equation without weff}), one gets a new differential equation:

\begin{equation}\label{equation sur constante rho}
    \frac{d}{dt}(\rho_{0,Hom})=T(\dot{s} +3Hs)a^{3(1+w)}.
\end{equation}

Using the chain rules to obtain derivatives with respect to the scale factor instead of time, one obtains the following expression for the density:

\begin{equation}\label{density as a function of a}
    \rho(a)=(\rho_0+\Delta \rho(a))a^{-3(1+w)},
\end{equation}

with

\begin{equation}\label{delta rho (a)}
    \Delta \rho(a) = \int_{a_0}^a\left( a' \frac{ds}{da'}(a') + 3 s(a')\right) T(a') {a'}^{2+3w} \,da'.
\end{equation}

Then, assuming a flat space, one can derive the relation between time and redshift using

\begin{equation}
    t-t_0=\int^{\frac{1}{1+z}}_{a_0\equiv 1} \frac{da}{aH} = \int^{\frac{1}{1+z}}_1 \frac{da}{a\sqrt{\frac{8 \pi G}{3} \rho}}. 
\end{equation}

By injecting Eq. (\ref{density as a function of a}), one gets

\begin{equation}\label{general time redshift relation with delta rho}
t-t_0= \frac{1}{H_0}\int^{\frac{1}{1+z}}_1 \frac{a^{\frac{1+3w}{2}}}{\sqrt{\Omega_0+\Delta \Omega (a)}} \; da,
\end{equation}

with $\Omega_0$ the usual cosmological density parameter, and 

\begin{equation}\label{delta omega(a)}
    \Delta \Omega (a) \equiv \frac{8 \pi G \Delta \rho(a)}{3H_0^2} \equiv \frac{\Delta \rho(a)}{\rho_{c,0}}.
\end{equation}

In the case of matter ($w=0$), and for constant temperature $T_0$, Eq. (\ref{general time redshift relation with delta rho}) can be expressed as

\begin{equation}\label{time and redshift relation for matter and T=cte}
t-t_0=\frac{1}{H_0} \int_1^{\frac{1}{1+z}} \frac{da}{\sqrt{(\Omega_{0}-\Omega_{S,0}) a^{-1}+\Omega_{S,0}\frac{s(a)}{s_0}a^2}} ,
\end{equation}

with $s_0$ the entropy density today and $\Omega_{S,0}=T_0 s_0/\rho_{c,0}$. 

\bibliography{refs.bib}

 \end{document}